\documentclass[iop]{emulateapj}

\shorttitle{Fullerenes and top-down chemistry}
\shortauthors{Zhen et al.}

\begin{document}

\title{Laboratory formation of fullerenes from PAHs:\\
  Top-down interstellar chemistry}

\author{Junfeng Zhen$^{1,2}$, Pablo Castellanos$^{1,2}$, Daniel M.\
  Paardekooper$^{2}$,\\Harold Linnartz$^{2}$, Alexander G.G.M.\
  Tielens$^{1}$} 

\affil{$^{1}$Leiden Observatory, Leiden University, P.O.\ Box 9513,
  2300 RA Leiden, The Netherlands}

\affil{$^{2}$Sackler Laboratory for Astrophysics, Leiden Observatory,
  Leiden University, P.O.\ Box 9513, 2300 RA Leiden, The Netherlands}

\email{zhen@strw.leidenuniv.nl}

\begin{abstract}
  Interstellar molecules are thought to build up in the shielded
  environment of molecular clouds or in the envelope of evolved
  stars. This follows many sequential reaction steps of atoms and
  simple molecules in the gas phase and/or on (icy) grain
  surfaces. However, these chemical routes are highly inefficient for
  larger species in the tenuous environment of space as many steps are
  involved and, indeed, models fail to explain the observed high
  abundances. This is definitely the case for the C$_{60}$ fullerene,
  recently identified as one of the most complex molecules in the
  interstellar medium. Observations have shown that, in some PDRs, its
  abundance increases close to strong UV-sources. In this letter we
  report laboratory findings in which C$_{60}$ formation can be
  explained by characterizing the photochemical evolution of large
  PAHs.  Sequential H losses lead to fully dehydrogenated PAHs and
  subsequent losses of C$_{2}$ units convert graphene into cages. Our
  results present for the first time experimental evidence that PAHs
  in excess of 60 C-atoms efficiently photo-isomerize to
  Buckminsterfullerene, C$_{60}$. These laboratory studies also attest
  to the importance of top-down synthesis routes for chemical
  complexity in space.
\end{abstract}


\section{Introduction}
\label{sec:intro}

Over the last 80 years, observational studies have revealed the
presence of $\sim$180 different molecules in space
\citep{her09,tie13}. Because of intrinsic limitations the
observational techniques, the molecular inventory revealed by these
studies is heavily biased towards small polar species, radicals, and
linear carbon chains with electronegative groups (e.g., CN). Also
larger (6--10 atoms containing) stable molecules like methanol and
ethylene glycol have been identified \citep{her09}. With the advent of
infrared space missions, a richer and diverser molecular universe was
brought to light \citep[and references therein]{tie08}. Most
mid-infrared spectra are dominated by broad features at 3.3, 6.2, 7.7,
8.6 and 11.2~$\mu$m, generally attributed to IR fluorescence of
UV-pumped, large (50--100 C-atoms) Polycyclic Aromatic Hydrocarbon
(PAHs) molecules \citep[and references therein]{tie08}. These
molecules contain $\sim$10\% of the elemental carbon and play an
important role in the ionization and energy balance of the
interstellar medium (ISM) of galaxies. Recently, in addition to these
PAH bands, the infrared signatures of Buckminsterfullerene, C$_{60}$,
were also observed at 7.0, 8.5, 17.4 and 18.9~$\mu$m
\citep{cam10,sel10}. PAH bands are also prominent in planet-forming
disks around young stars \citep{hab06,dou07} and PAHs as well as
C$_{60}$ are important components of solar system meteorites
\citep{sep08,bec94}. Hence, understanding the processes that regulate
the origin and evolution of these species and their relationship to
the organic inventory of space has become a focus in astrochemistry.

In the tenuous ISM, direct synthesis of PAHs and fullerenes from small
hydrocarbon species is inhibited \citep{mce99} and PAHs are generally
thought to form in the C-rich ejecta of Asymptotic Giant Branch stars
as molecular intermediaries or byproducts of the soot formation
process \citep{fre89,che92}. In the ISM, these species are then
processed by ultraviolet photons, which leads initially to their
ionization. Subsequently, UV photolysis results in dissociation, and
sequential steps of double H-losses, have been identified as the
dominant fragmentation channel, leading to pure carbon clusters,
likely in the form of graphene sheets
\citep{eke98,job03,ber12,zhe14}. After complete dehydrogenation,
ongoing photolysis will break down the carbon skeleton leading to
smaller carbon species. It has been suggested that highly excited
graphene sheets may also isomerize to more stable carbon cages or
fullerenes \citep{ber12}. Rapid transformation of graphene flakes into
the C$_{60}$ fullerene is observed in electron irradiation experiments
of graphene on surfaces \citep{chu10}. Fullerenes also fragment in a
strong UV field through the loss of C$_2$ units, shrinking the size of
the cage until the smallest photostable fullerene, C$_{32}$, is
reached \citep{han95}. From this point on, fragmentation leads to the
formation of rings and chains \citep{lif00}.

Photodissociation Regions (PDRs) provide a natural laboratory for the
study of the interaction of UV photons with carbonaceous species
\citep{pet05,rap05,hol99}. Observations with the \textit{Spitzer Space
  Telescope} and the \textit{Herschel Space Observatory} of the
prototypical PDR, NGC~7023, have revealed that the C$_{60}$ abundance
increases by an order of magnitude while the PAH abundance decreases
when approaching the illuminating star \citep{ber12}. These
observations point towards the important role of photochemistry in the
destruction of interstellar PAHs and that C$_{60}$ is likely a
photochemical product of PAHs \citep{ber12,cas14}. So far,
experimental evidence of this process has been lacking.

In this letter, we present laboratory results demonstrating the
formation of fullerenes (in particular C$_{60}$) from large PAHs by
photolysis, based on the difference in absorption properties as a
function of wavelength \citep{tat91}. We compare and contrast the
fragmentation pattern of fullerenes (C$_{60}$ and C$_{70}$) to that of
PAHs and their fragmentation products. As the C$_{60}$ fullerene does
not absorb at 532~nm, while PAHs and other fullerenes absorb
efficiently, the absorption behavior of C$_{60}$ fragments produced by
photolysis of PAHs can be used to establish the presence of C$_{60}$
fullerenes.

\section{Experimental Methods}
\label{sec:exp}

We have studied the fragmentation of fully-benzenoid PAH cations and
fullerene cations in the laboratory using i-PoP, our instrument for
Photodissociation of PAHs, which is described in detail in
\citet{zhe14}. Briefly, PAHs or fullerenes are sublimated in an oven,
at an appropriate temperature, ionized by an electron gun, and
transported into a quadrupole ion-trap via an ion gate. Later, the
cations are irradiated by many (typically $\sim$18) pulses from a
nanosecond pulsed Nd:YAG laser, leading to sequential steps of
fragmentation. The ion-trap content is subsequently released and
analyzed using a reflectron time-of-flight mass spectrometer. Each
fragmentation step is initiated by absorption of multiple photons; the
exact number depends on the laser wavelength. The process is heavily
biased towards dissociation through the lowest energy channel
\citep{zhe14}.

Here, we study the photo-fragmentation behavior of the large PAH
cations C$_{60}$H$_{22}^{\phantom{22}+}$ ($m/z =742.172$),
C$_{66}$H$_{26}^{\phantom{26}+}$ ($m/z = 818.203$), and
C$_{78}$H$_{26}^{\phantom{26}+}$ ($m/z = 962.203$). The mass spectra
are contrasted with those resulting from photo-fragmentation of the
fullerene cations C$_{60}^{\phantom{60}+}$ ($m/z = 720$) and
C$_{70}^{\phantom{70}+}$ ($m/z = 840$), in order to investigate the
formation of C$_{60}$ from large PAHs. These particular PAHs are
selected because their armchair edges provide them with greater
stability than PAHs with zigzag edges \citep{poa07,kos08}, which may
favor their presence in space. Indeed, PAHs with armchair edges are
observed to be more abundant in regions close to strong UV sources
\citep{can14}.

\section{Results}
\label{sec:results}

\begin{figure}[t]
  \centering
  \includegraphics[width=\columnwidth]{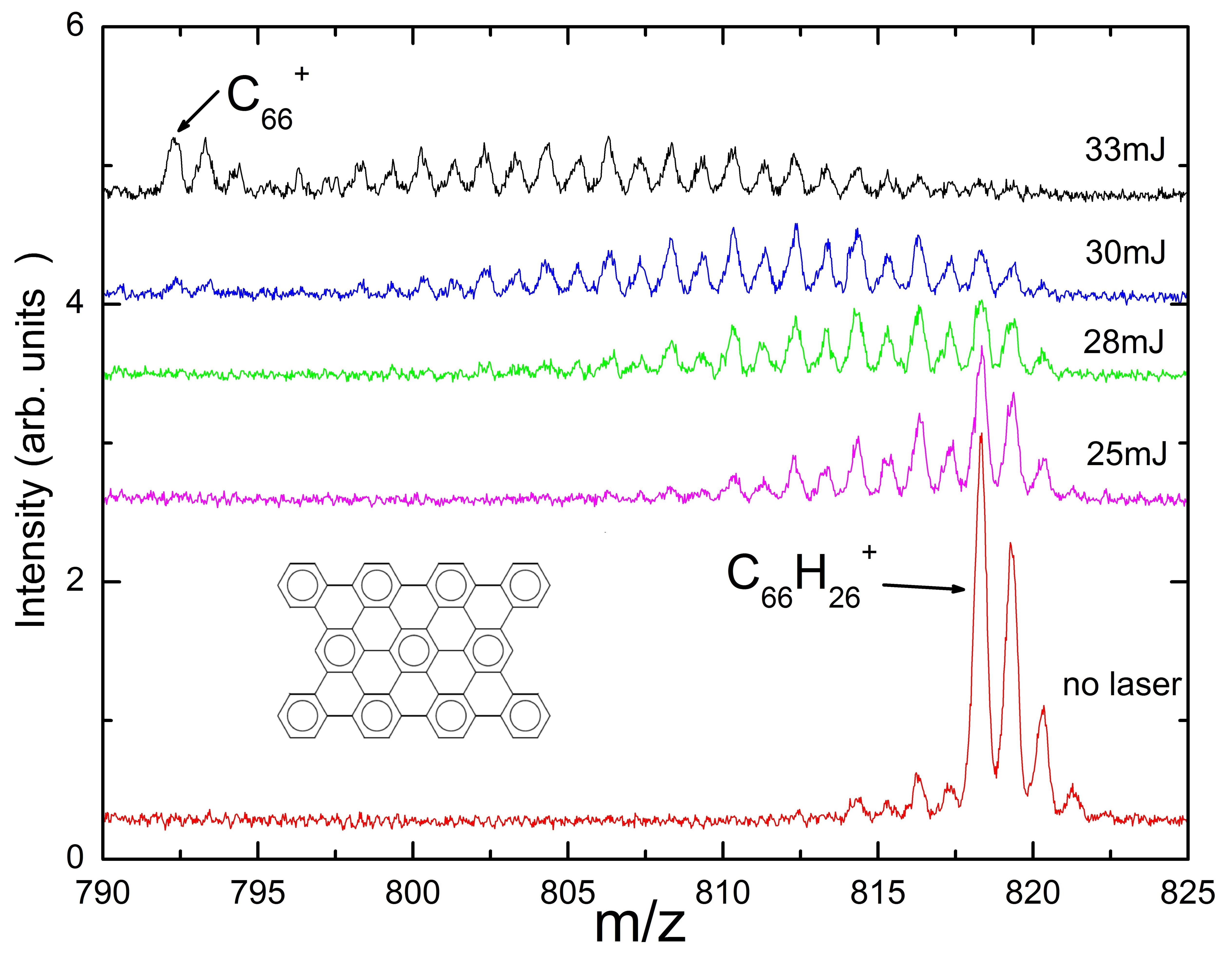}
  \caption{The fragmentation pattern of the fully benzenoid cation,
    C$_{66}$H$_{26}$$^{+}$, irradiated at 355~nm as a function of
    laser power. Additional peaks in the no-laser trace are isotopes
    and fragmentation produced by the electron gun.}
  \label{fig:dehyd}
\end{figure}

\begin{figure*}[t]
  \centering
  \includegraphics[width=\textwidth]{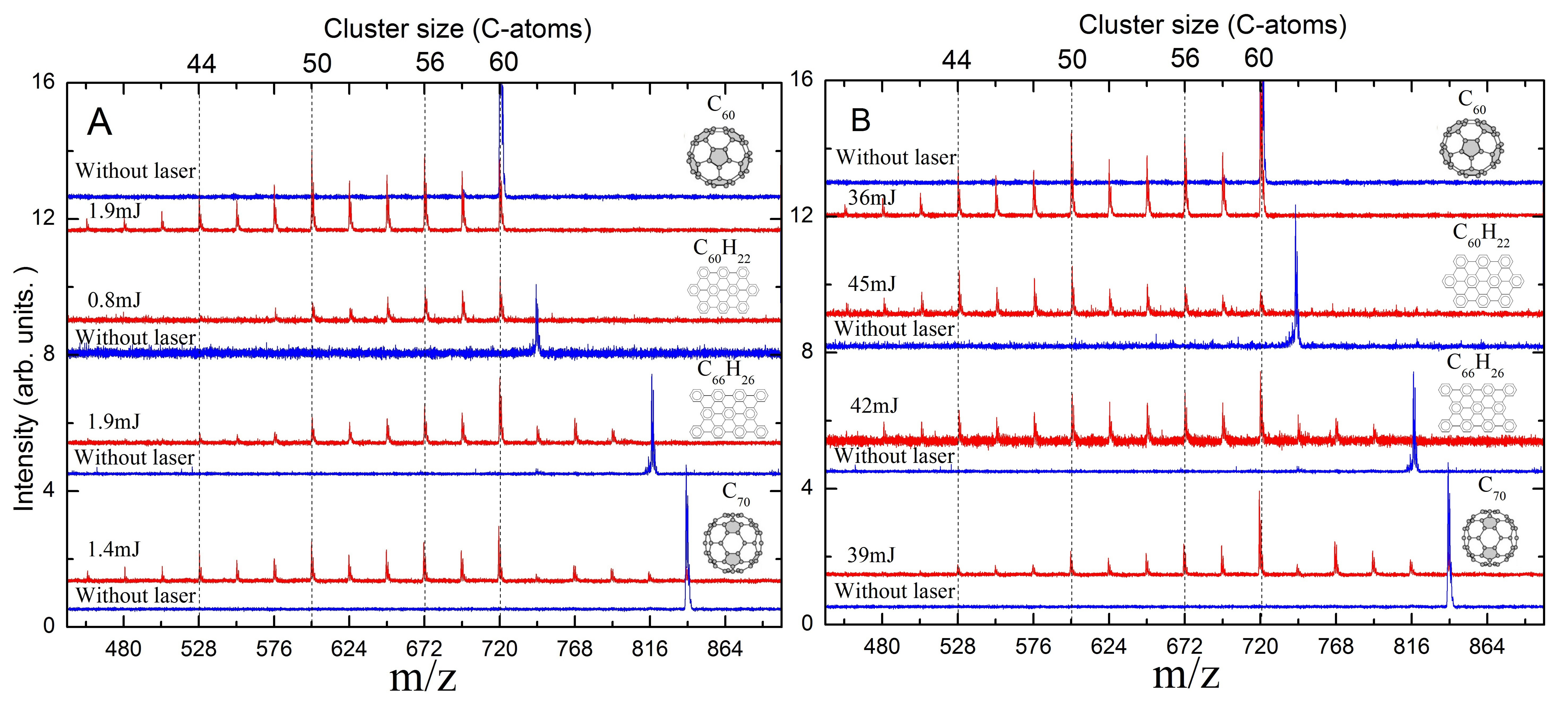}
  \caption{A comparison of the fragmentation pattern of the fully
    benzenoid cations, C$_{60}$H$_{22}$$^{+}$ and
    C$_{66}$H$_{26}$$^{+}$, and the fullerene cations, C$_{60}$$^{+}$
    and C$_{70}$$^{+}$, irradiated at 266 (panel A) and 355~nm (panel
    B). Note that the fragmentation pattern of the PAHs resembles that
    of the fullerenes. The ``magic numbers'' are marked by the
    vertical dashed lines and correspond to peaks with an $m/z$ of 720
    for C$_{60}$$^{+}$, 672 for C$_{56}$$^{+}$, 600 for C$_{50}$$^{+}$
    and 528 for C$_{44}$$^{+}$.}
  \label{fig:frag}
\end{figure*}

Studies at our laboratory and elsewhere have shown that, for small
PAHs, several fragmentation channels are available \citep[e.g., H
loss, C$_{2}$H$_{2}$ loss;][]{joc94}. However, for large PAHs,
fragmentation is almost exclusively through sequential hydrogen loss
\citep{eke98,job03,zhe14}. As an example, the dehydrogenation of
C$_{66}$H$_{26}^{\phantom{26}+}$ the hydrogenation state moves
progressively towards fully dehydrogenated species
(Figure~\ref{fig:dehyd}).  At even higher laser powers, these pure
carbon molecules and the fullerenes fragment through the loss of
C$_{2}$ units (Figure~\ref{fig:frag}), as has been observed in other
studies \citep{lif00}.

Fullerene formation in graphite vaporization experiments routinely
gives rise to ``magic numbers'' of C-atoms (60, 56, 50, 44) with
enhanced abundances compared to neighboring peaks, reflecting their
greater stability \citep{zim91}. We can use this pattern as an
indication of the formation of fullerene structures in our
photofragmentation studies. Panels A and B in Figure~\ref{fig:frag}
show the fragmentation through C$_{2}$ losses experienced by
C$_{60}^{\phantom{60}+}$, C$_{60}$H$_{22}^{\phantom{22}+}$,
C$_{66}$H$_{26}^{\phantom{26}+}$ and C$_{70}^{\phantom{70}+}$ at
266~nm and 355~nm, respectively. We observe that, in both cases, the
fullerene ``magic number'' mass peaks show an enhanced intensity with
respect to neighbouring peaks irrespective of the parent molecule
except for the C$_{60}^{\phantom{60}+}$ fragment from
C$_{60}$H$_{22}^{\phantom{22}+}$, where the enhancement is minimal
when compared to that of the other parent molecules. The differences
observed between the two wavelengths are due to variations in laser
power --- at shorter wavelengths, the laser system has a limited power
output ---, which is particularly clear for the PAHs, where the
smallest fragments are not produced at 266~nm.

Figure~\ref{fig:frag532} shows the C-losses for the same species at
532~nm. In this case the dissociation behavior of the fullerenes and
PAHs is markedly different. For the fullerenes, the dissociation stops
when reaching clusters with 60 C-atoms (in the case of
C$_{60}^{\phantom{60}+}$ no dissociation is observed). The PAHs
considered here experience fragmentation up to much smaller
masses. However, there is a striking difference in the behavior of the
C$_{60}^{\phantom{60}+}$ peak. For C$_{60}$H$_{22}^{\phantom{22}+}$
this peak does not appear to be special when compared to those further
down the line, while in the case of C$_{66}$H$_{26}^{\phantom{26}+}$
it shows a clear enhancement and remains even when all neighbouring
peaks have practically disappeared.

The kinetics of the fragmentation process is controlled by the
absorption properties of the parent species and its daughter products,
and by the dissociation energy of the fragmentation channels
involved. The dependence on absorption properties provides a tool with
which the fragmentation products can be probed. The PAH cations and
the fullerenes C$_{60}^{\phantom{60}+}$ and C$_{70}^{\phantom{70}+}$
as well as their products absorb well at 266 and 355~nm
\citep{tat91,mal07}. However, C$_{60}^{\phantom{60}+}$ does not absorb
at 532~nm \citep{tat91} and does not fragment, even at very high laser
powers (Figure~\ref{fig:frag532}). For C$_{70}^{\phantom{70}+}$, it is
clear that C$_{60}^{\phantom{60}+}$ is formed, thus halting the
fragmentation at 532~nm. The fragmentation pattern of
C$_{60}$H$_{22}^{\phantom{22}+}$ is not indicative for a hard to
dissociate 60 C-atom cluster, ruling out formation of significant
amounts of Buckminsterfullerene. Fragmentation of the pure carbon
cluster produced from C$_{66}$H$_{26}^{\phantom{26}+}$ can be
represented by a combination of the behavior of the fullerene
C$_{70}^{\phantom{70}+}$ and that of the
C$_{60}$H$_{22}^{\phantom{22}+}$ PAH cation, indicating formation of
the C$_{60}$ fullerene.

\begin{figure}[t]
  \centering
  \includegraphics[width=\columnwidth]{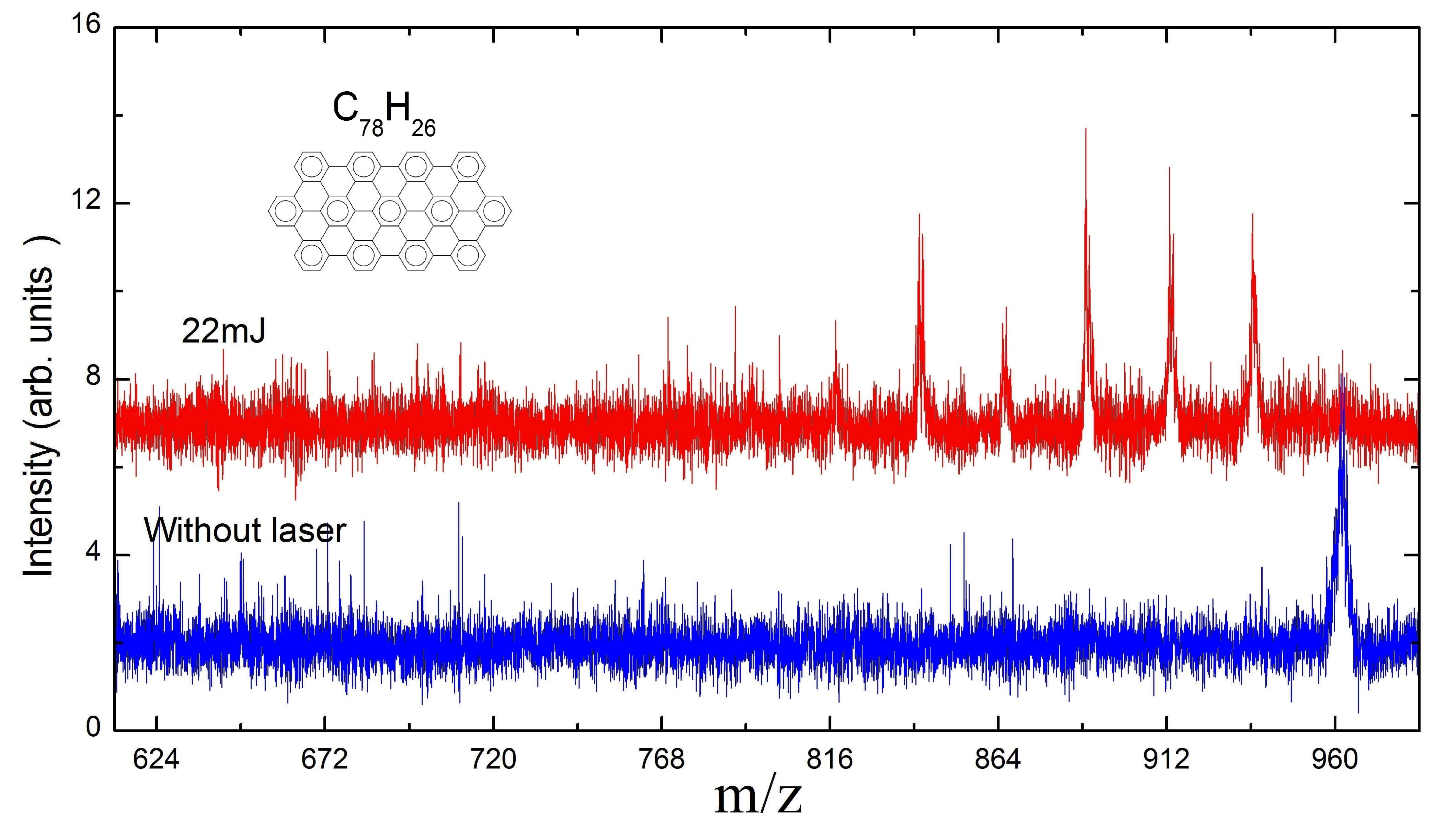}
  \caption{Fragmentation pattern of the fully benzenoid cation,
    C$_{78}$H$_{26}$$^{+}$, irradiated at 355~nm. Note that carbon
    clusters with 70 C atoms \textbf{($m/z = 840$)} show enhanced
    abundances. Due to signal-to-noise limitations this spectrum was
    taken at energies lower than those from Figure~\ref{fig:frag}. For
    this reason fragmentation to lower masses is not achieved.}
  \label{fig:C78}
\end{figure}

The formation of fullerenes from large PAHs may be a more general
process. The fragmentation pattern observed for
C$_{78}$H$_{26}^{\phantom{26}+}$ shows an enhanced intensity of the
peak corresponding to the C$_{70}^{\phantom{70}+}$ carbon cluster, as
illustrated in Figure~\ref{fig:C78}.

\section{Discussion}
\label{sec:disc}

The fragmentation pattern of PAHs and fullerenes --- specifically the
route through C$_{2}$ loss from C$_{60}$ --- can be quantified through
the C$_{58}^{\phantom{58}+}$/C$_{60}^{\phantom{60}+}$ ratio.
Figure~\ref{fig:ratio} demonstrates the similarity in behavior of the
C$_{60}^{\phantom{60}+}$ fragments produced from the PAHs and
fullerenes at 266 and 355~nm. However, at 532~nm, the
C$_{60}^{\phantom{60}+}$ fragments produced from
C$_{66}$H$_{26}^{\phantom{26}+}$ behave like those produced from
C$_{70}^{\phantom{70}+}$ (i.e.\ there is very little fragmentation
even at high laser power), while the C$_{60}^{\phantom{60}+}$
fragments produced from C$_{60}$H$_{22}^{\phantom{22}+}$ behave
similarly to the other wavelengths and fragment readily.

\begin{figure}[t]
  \centering
  \includegraphics[width=\columnwidth]{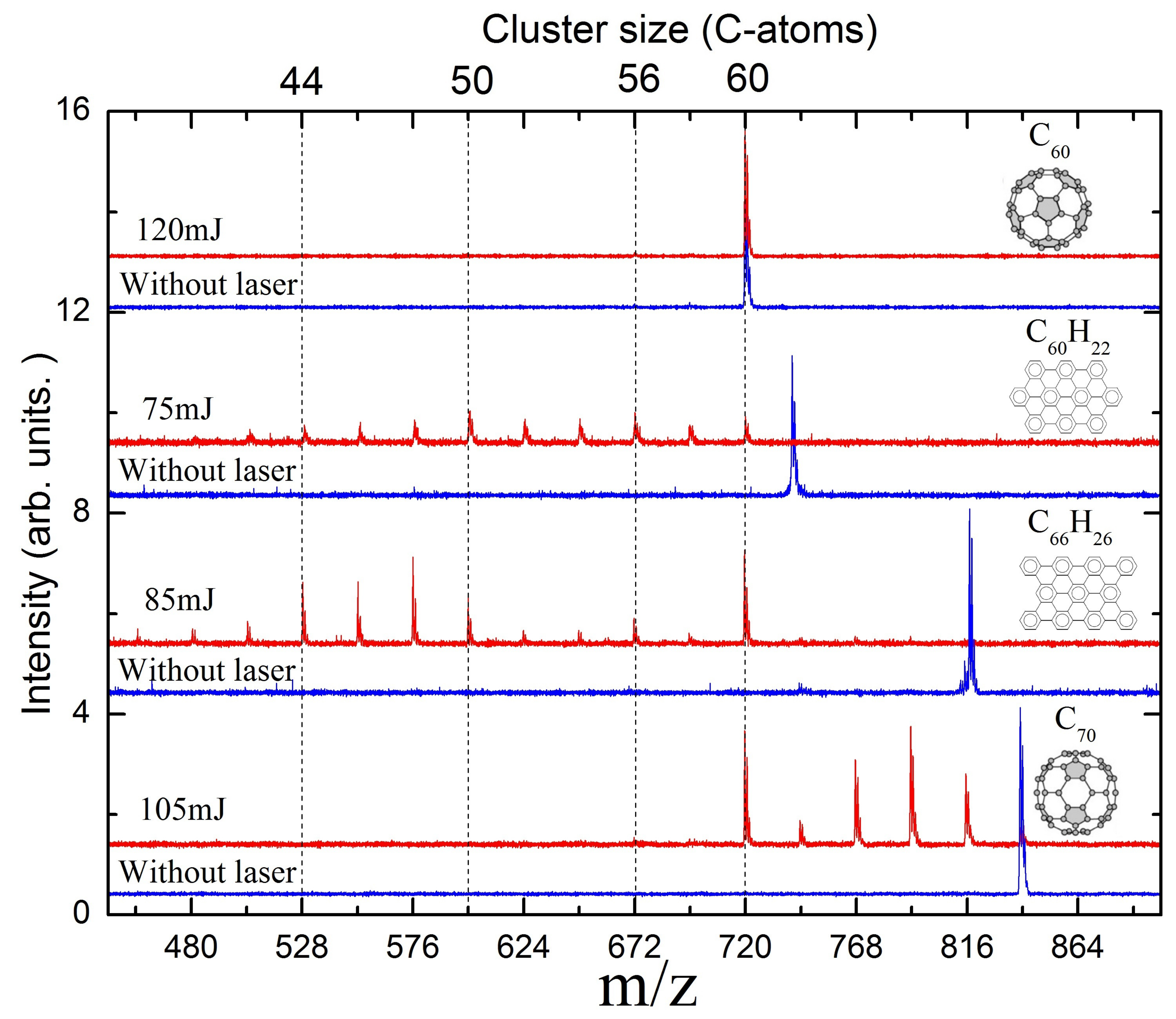}
  \caption{A comparison of the fragmentation pattern of the molecules
    considered in Figure~\ref{fig:frag}, irradiated at 532~nm. At this
    wavelength, the fullerene C$_{60}$$^{+}$ cation does not absorb
    and no fragmentation is observed. The C$_{60}$$^{+}$ species
    produced from C$_{60}$H$_{22}$$^{+}$ fragment readily to smaller
    clusters, while those produced from C$_{66}$H$_{26}$$^{+}$ show a
    mixed fragmentation behavior. As in Figure~\ref{fig:frag} the
    vertical dashed lines correspond to the fullerene ``magic
    numbers''.}
  \label{fig:frag532}
\end{figure}

Based on Figures~\ref{fig:frag532} and \ref{fig:ratio}, we conclude
that fragmentation of C$_{66}^{\phantom{66}+}$ leads to the presence
of both non-fullerene C$_{60}^{\phantom{60}+}$ isomer(s) as well as
the C$_{60}^{\phantom{60}+}$ fullerene. We conclude that the loss of
C$_{2}$ units from the pure carbon cluster, C$_{66}^{\phantom{66}+}$,
initiates isomerization of some of the initial fully dehydrogenated
PAHs to fullerene cages, which --- similar to the
C$_{70}^{\phantom{70}+}$ --- subsequently shrink to smaller and
smaller fullerene cages. The results show that this process is very
efficient, and a large fraction of the initial clusters ($\sim$20\% at
85~mJ) are channeled to the C$_{60}^{\phantom{60}+}$ fullerene. In
contrast, since the isomerization process is initiated by C$_{2}$
loss, the C$_{60}^{\phantom{60}+}$ fullerene formation channel is
essentially closed for the C$_{60}^{\phantom{60}+}$ produced by
dehydrogenating C$_{60}$H$_{22}^{\phantom{22}+}$ and this species
fragments fully to smaller species at 532~nm. We note though that the
fragmentation pattern of this PAH for products with $m/z < 720$
resembles that of the fullerenes C$_{60}$ and C$_{70}$, with peak
enhancements at the ``magic numbers''. Hence, after the
C$_{60}^{\phantom{60}+}$ formed from C$_{60}$H$_{20}^{\phantom{20}+}$
loses the first C$_{2}$, isomerization to cages can be initiated and
at this point further fragmentation follows the ``cage-route'' as
well.

Molecular dynamics calculations of the transformation of graphene
flakes to fullerenes have revealed that, at high temperatures, this
folding process starts through transformation of hexagons at the edges
of the flake to various polygons \citep{leb12}. Our experiments
suggest that the polygon formation associated with this folding
process can also be initiated by the loss of C$_{2}$ units from the
edges of the fully dehydrogenated PAHs and that, once the process is
started, it is ``self-sustaining'' and the chemical energy released
quickly drives the complete reconstruction of the dehydrogenated PAH
to a fullerene cage.

In summary, we conclude that photofragmentation of PAHs with more than
60 C-atoms leads to the formation of Buckminsterfullerene
(C$_{60}$). However, fullerenes are not the only products and isomers
with different absorption and stability properties are also formed.

\section{Astrophysical relevance}
\label{sec:astro}

In contrast to our laboratory studies, excitation of PAHs in PDRs is
due to single photon absorption. However, irrespective whether one or
multiple photons are involved, rapid intramolecular vibrational
redistribution will leave the species highly vibrationally excited
from which it relaxes either through fragmentation or IR
fluorescence. From our experiments it is not possible to derive the
activation energy for the different fragmentation channels and further
experiments are needed to confirm if these energies can be reached by
single photon absorption in the ISM. In the present study, we will
assume that this is the case and, hence, we directly apply our
experimental results to the photo-processing of PAHs in space.

Our results provide further insight in the evolution of PAHs in the
PDR associated with the reflection nebula NGC~7023. The processes
taking place in this nebula are representative for other similar
environments in space. In this region, winds from the young Herbig Be
star, HD~200775, have blown a cavity in the molecular cloud inside
which the star was formed \citep{fue98}. This cavity has broken open
to the surrounding ISM. The PAH abundance is observed to decrease ---
starting at $\sim$25$\arcsec$ from the star (some $20\arcsec$ inside
of the PDR front; well within the cavity) from about $7\times 10^{-2}$
of the elemental carbon to about $2\times 10^{-2}$ of the elemental
carbon at $10\arcsec$ from the star. The fullerene abundance increases
from about $10^{-5}$ at the PDR front to about $10^{-4}$ some
$10\arcsec$ from the star \citep{ber12}.

We can now interpret these observations in terms of the laboratory
results presented here and the model described by \citet{ber12} and
\citet{mon13}. The first step in the PAH destruction and fullerene
formation process is the loss of peripheral H. H-coverage of PAHs is a
balance between UV induced fragmentation and reactions with atomic H
and is controlled by the parameter $\gamma = G_{0}/n_{\mathrm{H}}$,
where $G_{0}$ is the intensity of the UV radiation field in terms of
the average interstellar radiation field \citep{hab68} and
$n_{\mathrm{H}}$ is the atomic hydrogen density in cm$^{-3}$
\citep{tie05,lep97}. For small $\gamma$, a given PAH species will be
fully hydrogenated while for large $\gamma$, it is fully
dehydrogenated and the transition between these two hydrogenation
states is very rapid. The critical value of $\gamma$ separating these
two cases is not well known and depends on the PAH size. According to
\citet{ber12}, values of $\gamma \approx 3$ and $\gamma \approx 100$
can be calculated for circumcoronene, C$_{54}$H$_{18}$, and
circumovalene, C$_{66}$H$_{20}$, respectively, based upon kinetic
parameters adopted from experimental studies on small PAHs
\citep{joc94,tie05}. \citet{mon13} on the other hand, adopting
slightly different values for the kinetic parameters, predict that for
circumcoronene 50\% of the species are fully dehydrogenated for
$\gamma \approx 4\times 10^{-2}$ while for circumovalene, this occurs
at $\gamma \approx 4\times 10^{-1}$. Further experimental studies for
large PAHs relevant for the ISM will have to settle this issue.

\begin{figure}[t]
  \centering
  \includegraphics[width=\columnwidth]{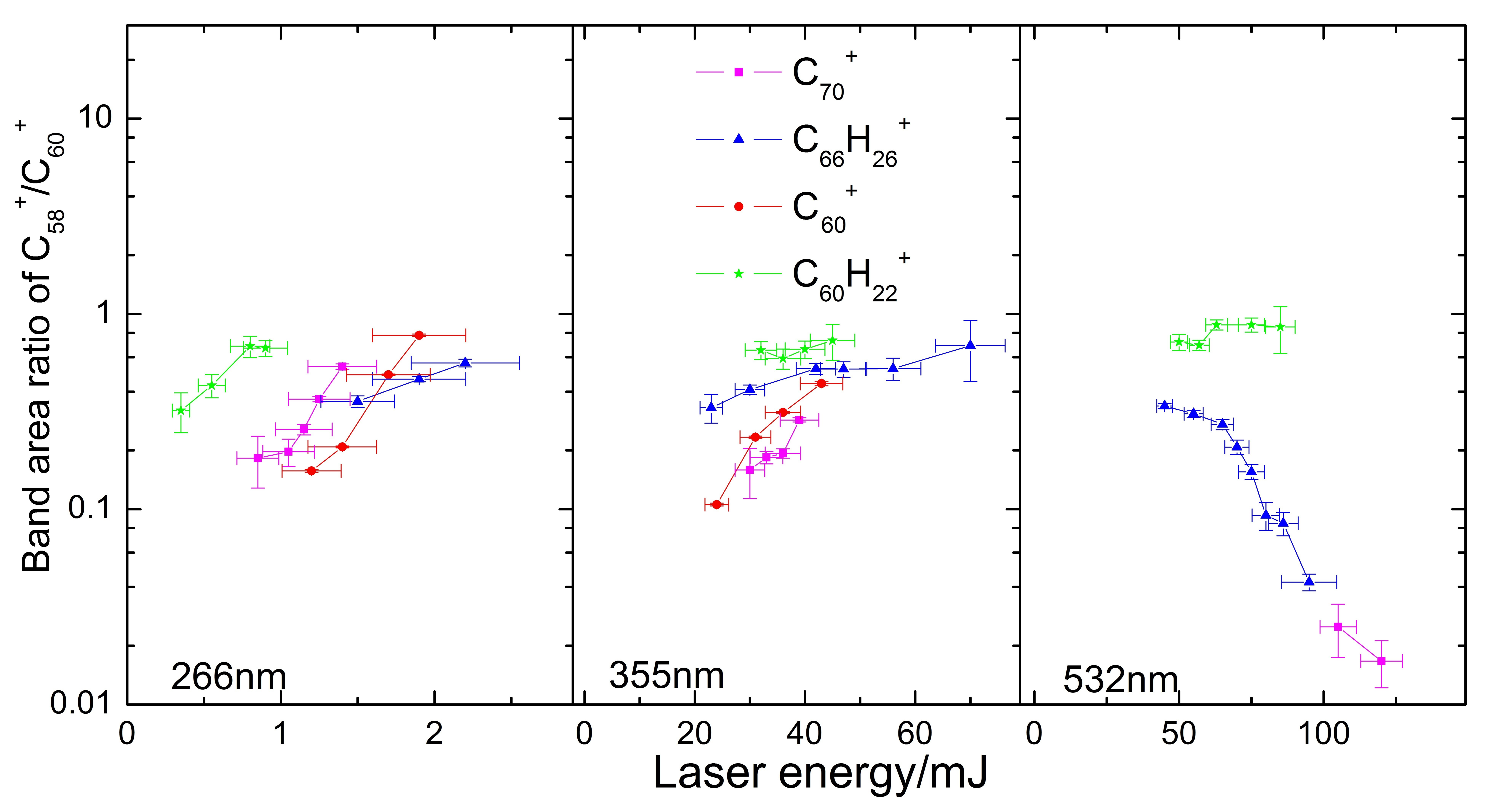}
  \caption{The ratio of the integrated intensity for
    C$_{58}$$^{+}$/C$_{60}$$^{+}$ as function of laser energy at
    different wavelengths --- 266~nm, 355~nm and 532~nm. For the first
    two wavelengths, the behavior of this ratio for PAH and fullerene
    cations is very similar. At 532~nm, the behavior of the ratio from
    C$_{60}$H$_{22}$$^{+}$ deviates from that of the other molecules
    considered.}
  \label{fig:ratio}
\end{figure}

We can compare these values to the $\gamma$ appropriate for the PDR in
NGC~7023. In the cavity of NGC~7023, between 12 and 25$\arcsec$ from
the star, where most of the PAH-to-fullerene conversion is observed to
occur, $\gamma$ is estimated to change from 65 to 650 \citep{ber12},
well into the regime where PAHs in excess of 60 C-atoms will become
dehydrogenated. Species that become fully dehydrogenated are rapidly
converted into cages but only those initially larger than 60 C-atoms
will form C$_{60}$. The high stability of C$_{60}$ allows it to
accumulate over irradiation time, as can be deduced from
Figure~\ref{fig:ratio}.

The observed high fraction of PAHs destroyed inside the cavity and the
relatively small amount of C$_{60}$ formed in NGC~7023 coupled with
the seemingly high efficiency with which large PAHs are converted into
the fullerene C$_{60}$ in our experiments suggests then that the
population of interstellar PAHs is heavily skewed towards PAHs smaller
than 60 C-atoms. The typical size of the interstellar PAH population
is not well known. In principle, the ratio of the intensity of the CH
stretching mode to the CH out-of-plane bending mode
($I_{3.3~\mu\mathrm{m}}/I_{11.2~\mu\mathrm{m}}$) is a measure of the
average size of the emitting PAHs \citep{all89,dra01,ric12}. However,
the infrared spectrometer on the Spitzer Space Telescope did not
extend to 3.3~$\mu$m while the beam of the Short Wavelength
Spectrometer on board of the Infrared Space Observatory encompasses
the full PDR in NGC~7023. Taking those latter values \citep{pee02} and
a typical excitation energy of 7~eV appropriate for a B3 star
($T_{\mathrm{eff}} = 17000$~K), the observed ratio (0.27) translates
into an average size of $\sim$60 C-atoms for the emitting PAHs in
NGC~7023 \citep{ric12}. Given the beam averaging in these
observations, this typical size is likely an
overestimate. Furthermore, comparisons of matrix isolation experiments
with DFT calculations suggest that the latter tends to overestimate
the intrinsic strength of the CH stretching mode by a factor of two
\citep{lan96}. Further observational and experimental studies are
required to address these issues.

\section{Conclusion}
\label{sec:concl}

In agreement with earlier studies, our photofragmentation studies
reveal that PAH cations initially fragment through rapid H-loss. The
resulting completely dehydrogenated species fragment further through
sequential steps of C$_{2}$ losses. We compared the results of the
fragmentation of the pure carbon clusters formed from PAH cations with
those of the fullerenes for different wavelengths.

Using the wavelength dependent absorption cross section properties of
C$_{60}^{\phantom{60}+}$, we demonstrate that PAH fragments formed
from PAHs that initially contain more than 60 C-atoms isomerize ---
among others species --- to the fullerene C$_{60}$. The presence of
``magic number'' peaks in the fragmentation pattern of smaller PAHs
suggests that these isomerize to small cages.

Based on our experimental studies, we have analyzed observational
results of PAHs and fullerenes in the NGC~7023 PDR, and conclude that
the observed PAH and C$_{60}$ relative abundances imply an
interstellar PAH size distribution skewed to sizes $\lesssim 60$
C-atoms. These experiments provide direct support for the importance
of top-down photochemistry for the formation of fullerenes in the ISM,
as well as other species. However, it must be noted that further
experiments are needed to determine the relevant fragmentation
energies and the feasibility that photons with these energies are
available in the ISM.

Finally, we recognize that, in a laboratory setting, the efficient
complete dehydrogenation of PAHs provides a novel way to synthesize
pure carbon clusters (most likely graphene flakes) of very specific
sizes. Given the present interest in graphene, this synthesis method
holds much potential for the study of such species under fully
controlled conditions and will allow a validation of theoretically
predicted properties by experiments.

\acknowledgments

We are grateful to M.J.A.\ Witlox and R.\ Koehler for technical
support. We are grateful to L.J.\ Allamandola who provided the large
PAH sample. Studies of interstellar chemistry at Leiden Observatory
are supported through advanced-ERC grant 246976 from the European
Research Council, through a grant by the Dutch Science Agency, NWO, as
part of the Dutch Astrochemistry Network, and through the Spinoza
premie from the Dutch Science Agency, NWO.

\end{document}